\documentclass[12pt]{article}
\usepackage{amsfonts}
\usepackage{amsmath}\allowdisplaybreaks[2]
\usepackage{amssymb}
\usepackage{mathptmx}
\usepackage{geometry}
\usepackage{graphicx}
\usepackage{hyperref}
\usepackage{cancel}
\usepackage{textcomp}
\usepackage{tikz}
\usepackage{bm}
\usepackage{mathrsfs}
\usepackage{filecontents}
\usepackage{times}
\usepackage{epsfig}
\usepackage{xcolor}
\usepackage{slashed}
\usepackage{latexsym}
\usepackage{verbatim}
\usepackage{extarrows}
\usepackage{cite}
\usepackage{multirow}
\usepackage{rotating}
\usepackage{colortbl}
\usepackage{indentfirst}
\usepackage{tensor}
\definecolor{mygray}{gray}{.9}
\definecolor{intnull}{RGB}{213,229,255}
\usepackage{arydshln}
\usepackage{diagbox}

\usepackage{caption}
\usepackage{tikz}
\usetikzlibrary{arrows,shapes,chains}
\usepackage{graphicx, subfig}
\begin{document}
\renewcommand{\thefootnote}{\fnsymbol{footnote}}
\baselineskip=16pt
\pagenumbering{arabic}
\vspace{1.0cm}
\begin{center}
{\Large\sf Absorption cross section of regular black holes in scalar-tensor conformal gravity}
\\[10pt]
\vspace{.5 cm}

{Yang Li\footnote{E-mail address: 2120190123@mail.nankai.edu.cn} and Yan-Gang Miao\footnote{Corresponding author. E-mail address: miaoyg@nankai.edu.cn}}
\vspace{3mm}

{School of Physics, Nankai University, Tianjin 300071, China}

\vspace{4.0ex}
\end{center}
\begin{center}
{\bf Abstract}
\end{center}

In terms of the complex angular momentum method, we compute the absorption cross section by analyzing a massless test scalar field around conformally related black holes. At first, we investigate circular null geodesics and thereby prove a precondition for calculating the absorption cross section in the context of conformally related black holes. Then we use the WKB approximation method to derive the analytic expression of Regge frequency and the oscillation part of absorption cross sections. We find that this oscillation part depends on the scale factor of conformal transformations. By taking  the conformally related Schwarzschild-Tangherlini black hole as an example, we show that this regular black hole has substantially distinctive absorption behavior compared with singular black holes. Our result provides a new approach to distinguish a regular black hole from a singular one.

\section{Introduction}
\label{sec:intr}

Searching for a gravitational theory beyond General Relativity (GR) has never stopped since GR was realized to be imperfect as a theory of gravitation. A serious defect is that the black hole (BH) spacetimes in GR include singularities, which is not an honest reflection of astrophysical objects. Many gravitational theories have been proposed to solve this problem, among which the conformal gravity is a promising attempt~\cite{Weyl,andp,ST}. Among various kinds of conformal gravitational theories that have been constructed so far, we focus on the conformal gravity from the scalar-tensor theory~\cite{1605,1611}.  The reason is that some BH solutions of GR belong to the family of solutions of the scalar-tensor conformal gravitational theory.  This is quite interesting. When a gravitational theory has a conformal invariance, one can obtain its metric by multiplying the metric of GR by a scale factor. As a consequence, the metric of the scalar-tensor conformal gravitational theory is composed of a GR metric multiplied by a scale factor. The black hole family obtained in this way is called conformally related black holes (CR BHs) and is usually a kind of regular black holes if a suitable scale factor is chosen.

The influence of scale factors is nontrivial. Significantly, the singularity located at the center of a BH can be removed~\cite{1611,2102} with a proper choice of scale factors, namely, CR BHs are usually regular. Besides the removing of spacetime singularities, various properties of CR BHs have been studied, including the quasinormal modes (QNMs) and dynamic stability~\cite{2102,1705, Pisin}, thermodynamics and phase transitions~\cite{1611thermo,1711,2102}, and the violation of energy conditions in CR BH spacetimes~\cite{1702}.

At present, the absorption property of CR BHs remains to be investigated. The absorption cross section (ACS) of a BH describes the interaction between the BH and the matter around it, which reflects the essential property of the BH. Some previous studies show~\cite{1408,1505,AP393,IJMPD26,2011} that the ACSs of regular BHs associated with nonlinear electrodynamics (NED) are not very different from those of the Reissner-Nordstr\"om BH. With fine tuning of parameters, the ACSs of the two kinds of BHs can even be the same. Because the Reissner-Nordstr\"om BH is singular, this phenomenon turns off the feasibility to distinguish the regular BHs associated with NED from singular ones by their ACSs. Nevertheless, the studies about the QNMs of CR BHs are inspiring, which indicates that this dynamic property is greatly distinct from that of singular BHs, depending on the choice of a scale factor. Therefore, it is of considerable value to investigate the ACSs of CR BHs from the perspective on distinguishing a regular BH from a singular one.

We adopt the complex angular momentum (CAM) method to derive the analytic expression of ACSs. The CAM method was originally applied to scattering problems in quantum mechanics. The pioneering progress was made by Watson~\cite{Watson}, see the monographs by Newton~\cite{Newton} and Nussenzveig~\cite{Nussenzveig} for the details. The first application of the CAM method to a gravitational theory was accomplished by Chandrasekhar and Ferrari~\cite{Chandrasekhar}, followed by Andersson and Thylwe~\cite{Andersson1,Andersson2} who applied this technique to the Schwarzschild BH. Later, the connection between surface waves and QNMs was established~\cite{0212093} in terms of the CAM method, where  the QNMs were regarded as the resonance modes of surface waves. Moreover, the related progress was  made~\cite{0906,1002,1101,1104,1108,Kerr}, which provided a general proposal to derive the analytic expression of Regge frequency and thereby to work out ACSs. Incidentally, one  method rather than the CAM was once proposed~\cite{Dolan} to calculate the Regge frequency. In this paper, we follow the way suggested in Refs.~\cite{0212093,0906,1002,1101,1104,1108,Kerr} but generalize it for dealing with CR BHs.

The outline of this paper is as follows. We briefly review the scalar-tensor conformal gravity in Sec.~\ref{sec:comformal gravtity} and then the BH scattering/absorption theory in Sec.~\ref{sec:review of scattering/absorption}. In Sec.~\ref{sec:geo}, we prove a precondition for calculating Regge frequency in the context of CR BHs.
We derive in Sec.~\ref{sec:Regge} the analytic expression of Regge frequency by using the third-order WKB method and apply the formula to the models of conformally related Schwarzschild-Tangherlini black holes (CRST BHs). In Sec.~\ref{sec:ACS}, we give the general expression of ACSs and show its dependence on scale factors.  We illustrate our results by taking the models of CRST BHs as examples. Finally, we present our conclusions in Sec.~\ref{sec:Conclusion}.

\section{General formalism}
\label{sec:general}

\subsection{Conformal gravity}
\label{sec:comformal gravtity}
The conformal gravitational theory we consider is described~\cite{1605,1611,1705} by the action,
\begin{equation}
	I_{\rm C}=\frac{1}{2}\int \mathrm{d}^{D}x \,\sqrt{-{g}} \phi\left(\frac{1}{4}\frac{D-2}{D-1}{R} \phi-{\square} \phi\right), 
	\label{confact}
\end{equation}
where $\phi$ is a massless scalar field, $R$ the Ricci scalar, and
$\square \equiv g_{\mu\nu}\nabla_{\mu}\nabla_{\nu}$ the covariant d'Alembertian. Eq.~(\ref{confact}) is invariant under the conformal transformations of $g_{\mu\nu}$ and $\phi$ as follows:
\begin{eqnarray}
	g_{\mu\nu}  &\longrightarrow&   S(x) g_{\mu\nu},\label{hatmetrictrans}\\
	\phi   &\longrightarrow&   [S(x)]^{(2-D)/4}\,\phi,\label{phitrans}
\end{eqnarray}
where $S(x)$ denotes the scale factor.

There are many solutions of action Eq.~(\ref{confact}). In general, $g_{\mu\nu}$ and $\phi$ depend on time. Here we focus on static solutions. In particular, if we choose $\phi=2\sqrt{\frac{D-1}{D-2}}$, the conformal gravity reduces to the Einstein gravity. Therefore, there is a family of solutions in conformal gravity Eq.~(\ref{confact}). This family of solutions is conformally related to the specific solution in the Einstein gravity through the conformal transformations Eqs.~(\ref{hatmetrictrans}) and (\ref{phitrans}). Based on the metric of a spherically symmetric BH in the Einstein gravity, one can write~\cite{1605,1611,1705} the metric of the family of conformally related BHs,
\begin{equation}
	\mathrm{d}s^2=S(r)\left[-f(r)\mathrm{d}t^{2}+\frac{1}{f(r)}\mathrm{d}r^{2}+r^{2}\mathrm{d}\Omega_{D-2}\right],
	\label{metric2}
\end{equation}
where $f(r)$ is the lapse function in the Einstein gravity.
Eq.~(\ref{metric2}) is asymptotically flat in the $D$-dimensional spacetime because both $S(r)$ and $f(r)$ equal one in the limit of $r\to \infty$.
With special choices of the scale factor, the spacetime singularity of Eq.~(\ref{metric2}) can be removed, which can be verified~\cite{1605,2102,Pisin} by finite curvature invariants and complete geodesics in the spacetime.

\subsection{Review of scattering/absorption theory}
\label{sec:review of scattering/absorption}

To study the dynamics of spacetime Eq.~(\ref{metric2}), one can investigate the simplest case: the minimally coupled massless scalar field. The dynamics of this test scalar field is described by the Klein-Gordon equation,
\begin{equation}
\frac{1}{\sqrt{-g}}\partial_{\mu}\sqrt{-g}g^{\mu\nu}\partial_{\nu}\Phi=0.
\label{KG eq.}
\end{equation}
The decomposition of $\Phi$ in the background of Eq.~(\ref{metric2}) has been introduced~\cite{2102},
\begin{equation}
\Phi=\sum_{\ell,m} \frac{1}{r^{(D-2)/2} [S(r)]^{(D-2)/4}}e^{-i \omega t}\psi_{\ell}(r)\mathrm{Y}_{\ell m}
(\theta_{1}, \ldots, \theta_{D-2}),
\label{decomposition}
\end{equation}
where $\mathrm{Y}_{\ell m}(\theta_{1}, \ldots, \theta_{D-2})$ stands for the spherical harmonics of $D-2$ angular coordinates, $0\leq \left( \theta_{1}, \theta_{2}, \ldots, \right.$ $\left. \theta_{D-3} \right) \leq \pi $, and $0 \leq \theta_{D-2} \leq 2 \pi$.
With the help of Eqs.~(\ref{KG eq.}) and (\ref{decomposition}), one can obtain the Schr\"odinger-like equation,
\begin{equation}
\partial_{r_{*}}^2 \psi_{\ell}+\left( \omega^{2}-V_{\ell} \right)\psi_{\ell}=0,
\label{master eq}
\end{equation}
with  the effective potential,
\begin{equation}
V_{\ell}=f(r)\left\{ \frac{\ell(\ell+D-3)}{r^2} + \frac{\left( f(r)\left(r^{(D-2)/2}[S(r)]^{(D-2)/4}\right)^{\prime} \right)^{\prime}}{[S(r)]^{(D-2)/4}r^{(D-2)/2}} \right\},
\label{V_l}
\end{equation}
where the ``tortoise" coordinate $r_{*}$ is defined as $\mathrm{d} r_{*}\equiv\mathrm{d}r / f(r)$ and the prime means the derivative with respect to the radial coordinate.
The scalar wave is purely ingoing at the event horizon and outgoing at the spatial infinity.
Under these boundary conditions, the eigenvalues $\omega$ of the master equation Eq.~(\ref{master eq}) are discrete. Therefore,  one obtains~\cite{Berti,Kokkotas,Nollert,KonoplyaQNM} eigenmodes, {\em i.e.} the QNMs as well as the complex quasinormal frequencies (QNFs), $\omega=\omega_{\ell n}-i\Gamma_{\ell n}$,
where $\omega_{\ell n}$ represents the oscillation and $\Gamma_{\ell n}$ the inverse of a damping time scale.

The dynamics of the test scalar field can also be understood in the perspective of BH scattering.
The boundary conditions can be written explicitly as follows:
\begin{itemize}
\item At the event horizon, the partial wave $\psi_{\ell}(r)$ is purely ingoing without reflection,
\begin{equation}
\psi_{\ell}(r) \sim e^{-i\omega r_{*}}, \qquad r_{*} \rightarrow - \infty;
\label{at horizon}
\end{equation}
\item At the spatial infinity, $\psi_{\ell}(r)$ is asymptotically free,
\begin{equation}
\psi_{\ell}(r) \sim \frac{1}{T_{\ell}(\omega)}e^{-i\omega r_{*}+i(\ell+\frac{D-3}{2})\frac{\pi}{2}-i\frac{\pi}{4}} - \frac{S_{\ell}(\omega)}{T_{\ell}(\omega)}e^{+i\omega r_{*}-i(\ell+\frac{D-3}{2})\frac{\pi}{2}+i\frac{\pi}{4}}, \qquad r_{*} \rightarrow + \infty.
\label{at infinity}
\end{equation}
Note that  $1/T_{\ell}(\omega)$ is the reflective coefficient, $S_{\ell}(\omega)$ is the element of diagonalized scattering matrix, and $S_{\ell}(\omega)/T_{\ell}(\omega)$ is the transmitted coefficient. The terms $\pm i(\ell+\frac{D-3}{2})\frac{\pi}{2}\mp i\frac{\pi}{4}$ on the exponentials are relative phase differences between the reflective and transmitted waves.
\end{itemize}
According to the boundary conditions, the scalar wave is asymptotically free at the spatial infinity, and thus it does not encounter any potential barriers. Therefore, the reflective coefficient in Eq.~(\ref{at infinity}) should vanish, while the transmitted coefficient should survive. This implies that the boundary conditions pick the poles of the scattering matrix, at which both $T_{\ell}(\omega)$ and $S_{\ell}(\omega)$ are infinite, but the ratio $S_{\ell}(\omega) / T_{\ell}(\omega)$ is finite.
As $S_{\ell}(\omega)$ is a function of both $\ell$ and $\omega$, the poles can be interpreted in two perspectives.

In the first perspective, $S_{\ell}(\omega)$ is analytically continued to the complex $\omega$-plane ($\omega \in \mathbb{C}$), while $\ell \in \mathbb{N}$ remains unchanged. In this case, the poles of the scattering matrix correspond to the QNFs. As a consequence, on the complex-$\omega$ plane, the scattering matrix $S_{\ell}(\omega)$ can be expanded in the vicinity of QNFs,
\begin{equation}
S_{\ell}(\omega) \propto \frac{\Gamma_{\ell n}}{2(\omega-\omega_{\ell n}+i\Gamma_{\ell n})},
\end{equation}
which is the Breit-Wigner type of resonances; see Appendix A of Ref.~\cite{1002} for the details.

In the second perspective we adopt in this paper, $S_{\ell}(\omega)$ is analytically continued to the complex $\lambda$-plane (CAM plane) by defining $\lambda \equiv \ell+(D-3)/2$ at first and then analytically continuing $\lambda$ to be complex, while $\omega \in \mathbb{R}$ remains unchanged. In this case, the poles of the scattering matrix correspond to the Regge poles. Conventionally, the notation $\lambda_n(\omega)$, $n=1,2,\ldots$, is used to distinguish different Regge poles.

As suggested in Refs.~\cite{0212093,0906,1002,1101,1104,1108,Kerr}, one can interpret resonances as the surface waves travelling around a photon sphere and damping simultaneously. $\mathrm{Re}\,\lambda_n(\omega)$ denotes the speed of circumnavigation of surface waves and $\mathrm{Im}\,\lambda_n(\omega)$ the damping of surface waves. In fact, the QNFs are closely related to the Regge poles. When the resonances happen, i.e., the real part of Regge poles satisfies
\begin{equation}
\mathrm{Re} \mspace{3mu}\lambda_n(\omega_{\ell n}) = \ell + \frac{D-3}{2}, \qquad \ell \in \mathbb{N},
\label{Regge-Re(QNM)}
\end{equation}
one can solve the real part of QNFs. In addition, the minus imaginary part of QNFs can be determined by the Regge poles,
\begin{equation}
\Gamma_{\ell n}=\left.\frac{\mathrm{Im}\, \lambda_n(\omega)\,\frac{\mathrm{d}}{\mathrm{d}\omega}\, \mathrm{Re}\,\lambda_n(\omega)}{\left[\frac{\mathrm{d}}{\mathrm{d}\omega}\, \mathrm{Re}\,\lambda_n(\omega)\right]^2+\left[\frac{\mathrm{d}}{\mathrm{d}\omega}\, \mathrm{Im}\,\lambda_n(\omega)\right]^2}\right|_{\omega=\omega_{\ell n}}.
\label{Regge-Im(QNM)}
\end{equation}

\section{Proof of a precondition for conformally related black holes}
\label{sec:geo}

In this section, our main purpose is to prove a precondition used in the calculation of Regge frequency for  CR BHs.

The location of a photon sphere of a static spherically symmetric BH is determined~\cite{Chandra_BH,0812} by the following formulas  for the trivial scale factor,  $S(r)=1$,
\begin{eqnarray}
&&f'(r_c)-\frac{2}{r_c}f(r_c)=0,\label{PS1}\\
&&f''(r_c)-\frac{2}{r_c^2}f(r_c) < 0.
\label{PS2}
\end{eqnarray}
In fact, according to Ref.~\cite{1002}, one can approximatively obtain $r_0 \approx r_c$ in the large-$\ell$ limit, where $r_0$ stands for  the peak  of the effective potential $V_{\ell}$ (Eq.~(\ref{V_l})).
We need this approximate result in CR BHs with the nontrivial $S(r)$. Our way to prove is straightforward: If  $r_c$ and $r_0$ are independent of $S(r)$  in the large-$\ell$ limit, the approximate result holds for CR BHs. This is what we call the  precondition.

\subsection{$r_c$ independent of $S(r)$}
We start with the line element Eq.~(\ref{metric2}) and restrict our discussion, without loss of generality, to the circular geodesic orbits within the equatorial plane. As shown by Chandrasekhar~\cite{Chandra_BH} and Cardoso~\cite{0812}, the Lagrangian of a particle moving along one of such obits is\footnote{Due to the different sign conventions, Eq.~(\ref{null Lagrangian}) is different from the Lagrangian in Ref.~\cite{0812} by a minus sign.}
\begin{equation}
2\mathcal{L}=-S(r)f(r)\dot{t}^2+\frac{S(r)}{f(r)}\dot{r}^2+S(r)r^2\dot{\varphi}^2,
\label{null Lagrangian}
\end{equation}
where the dot denotes the derivative with respect to the affine parameter $\tau$ and $\varphi$ stands for the azimuthal angular coordinate. The generalized momenta then take the forms,
\begin{subequations}
\begin{eqnarray}
p_t &=& -S(r)f(r)\dot{t}\equiv -E,\\
\label{momentum1_t}
p_{\varphi} &=& S(r)r^2\dot{\varphi} \equiv L,\\
\label{momentum1_phi}
p_r&=&\frac{S(r)}{f(r)}\dot{r},
\label{momentum1_r}
\end{eqnarray}
\label{momentum1}
\end{subequations}
where $E$ and $L$ are conserved energy and angular momentum of the moving particle, respectively.
Solving Eq.~(\ref{momentum1}) for $\dot{t}$ and $\dot{\varphi}$, we obtain
\begin{eqnarray}
\dot{t} &=& \frac{E}{S(r)f(r)},\\
\label{dot_t}
\dot{\varphi} &=& \frac{L}{S(r)r^2},
\label{dot_phi}
\end{eqnarray}
and thus give the conserved Hamiltonian,
\begin{eqnarray}
2 \mathcal{H} &=& 2(p_t\dot{t}+p_{\varphi}\dot{\varphi}+p_r\dot{r}-\mathcal{L})\nonumber\\
&=&-S(r)f(r)\dot{t}^2+\frac{S(r)}{f(r)}\dot{r}^2+S(r)r^2\dot{\varphi}^2
\nonumber\\
&=& -\frac{E^2}{S(r)f(r)}+\frac{S(r)}{f(r)}\dot{r}^2+\frac{L^2}{S(r)r^2}
\nonumber\\
&\equiv & \delta_1,
\label{Hamiltonian}
\end{eqnarray}
where $\delta_1=0$ stands for null geodesics, while $\delta_1=-1$ for time-like ones. Eq.~(\ref{Hamiltonian}) can be recast into
\begin{equation}
\dot{r}^2=V_r,
\label{r_dot}
\end{equation}
with
\begin{equation}
V_r= \frac{f(r)}{S(r)}\left[ \delta_1-\frac{L^2}{S(r)r^2}+\frac{E^2}{S(r)f(r)} \right],
\end{equation}
where $V_r$ denotes the geometric potential of particles. Note to distinguish $V_r$ from the effective scattering potential $V_{\ell}$.

The photon sphere consists of circular null geodesics at $r=r_c$, and these null geodesics are unstable orbits such that
\begin{equation}
\left.V_r \right|_{r=r_c}=0=\left.V_r'\right|_{r=r_c},
\label{Vr1}
\end{equation}
and
\begin{equation}
\left.V_r''\right|_{r=r_c}<0.
\label{Vr2}
\end{equation}
Eq.~(\ref{Vr1}) determines the value of $r_c$, while Eq.~(\ref{Vr2}) shows the instability of orbits. We can write Eq.~(\ref{Vr1}) explicitly,
\begin{equation}
\left.V_r \right|_{r=r_c}=0  \mspace{20mu}\Longrightarrow  \mspace{20mu}\frac{E^2}{f_c}-\frac{L^2}{r_c^2}=0,
\label{Vr3}
\end{equation}
and
\begin{equation}
\left.V_r'\right|_{r=r_c}=0 \mspace{20mu} \Longrightarrow \mspace{20mu} \left.\left( \frac{E^2}{\left[S(r)\right]^2}-\frac{L^2f(r)}{r^2S(r)}\right)'\right|_{r=r_c}=0,
\label{Vr4}
\end{equation}
where $f_c$ means $f(r_c)$. The similar notation is also applied to the scale factor, {\em i.e.}, $S_c\equiv S(r_c)$, $S_c^{\prime}\equiv S^{\prime}(r_c)$, {\em etc.}, in the following contexts.
Eq.~(\ref{Vr4}) yields
\begin{equation}
2S'_c r_c\left(f_c L^2-E^2 r_c^2\right)-L^2S_c\left(2f_c-r_c f'_c\right)=0.
\label{Vr5}
\end{equation}
Note that none of $r_c$, $S_c$, $S'_c$, and $L$ equal zero. Substituting Eq.~(\ref{Vr3}) into Eq.~(\ref{Vr5}), we exactly obtain Eq.~(\ref{PS1}) which is associated with the trivial scale factor, $S(r)=1$.
Hence, we prove that $r_c$ is independent of $S(r)$. In general, it is reasonable that the scale factor does not alter the location of photon spheres, since the equation of motion of null geodesics is conformally invariant.

\subsection{$r_0$ independent of $S(r)$}
Now we analyze $V_{\ell}$ for $\ell \gg 1$. $V_{\ell}$ has the following behavior in large-$\ell$ regimes for the CR BHs described by Eq.~(\ref{metric2}),
\begin{equation}
V_{\ell}  \sim  \frac{\ell^2f(r)}{r^2}, \qquad \ell \gg 1,
\label{V_l l_large}
\end{equation}
which can simply be verified when $\ell \gg 1$ is applied  to Eq.~(\ref{V_l}) and the terms proportional to $\ell^0$ and $\ell^1$ are omitted. Note that Eq.~(\ref{V_l l_large}) is exactly same as Eq.~(44) of Ref.~\cite{0812}. Therefore,  the peak $r_0$ of $V_{\ell}$ in large-$\ell$ regimes is also independent of $S(r)$.

As a summary of Section~\ref{sec:geo}, the approximate result, $r_0 \approx r_c$ for $\ell \gg 1$, still holds for CR BHs with the nontrivial $S(r)$ because both $r_0$ and $r_c$ are not altered by $S(r)$ in the large-$\ell$ limit. Incidentally, this implies that the geometric ACS, which will be discussed in detail in Section~\ref{sec:ACS}, is independent of $S(r)$.

\section{Regge frequency of conformally related static spherically symmetric black holes: WKB method}
\label{sec:Regge}

In this section we calculate the Regge frequencies for CR BHs in terms of the third-order WKB method, which is the basis for us to investigate ACSs in the next section.

\subsection{Regge frequency: General expression}
\label{sec:Regge_general}

Now we derive the analytic expression of Regge frequency via the third-order WKB method. Originally, the WKB method was used to derive the eigenvalues of Schr\"odinger equation in quantum mechanics; see Ref.~\cite{Sakurai}  for a pedagogical introduction. Then it was introduced into BH perturbation theory for the calculation of QNMs, because the master equations of perturbative fields have the Schr\"odinger-like form and the QNMs correspond to the eigenvalues of  Schr\"odinger-like equations. The first-order WKB method was applied by Schutz and Will~\cite{1st}, the third-order one by Iyer and Will~\cite{3rd}, the sixth-order one by Konoplya~\cite{6th}, and the higher-order one by Konoplya {\em et al.}~\cite{Higher,Pade}. Recently, the WKB method has also been used to calculate the Regge frequency in BH perturbation theory; see Ref.~\cite{0212093,0906,1002} for the details.

Here we resort to the third-order WKB method. As pointed out in Ref.~\cite{1002}, although the higher-order WKB method has higher precision, the complexity of calculations also grows greatly, which gives rise to a too complicated analytic expression. As a primary application to CR BHs, the third-order WKB is good enough, which can be seen in the following discussions.

We start with the third-order WKB equation~\cite{3rd} in which the QNFs satisfy
\begin{equation}
\omega^2=\left[V_0(\lambda)+\left[-2V_0^{(2)}(\lambda)\right]^{1/2}\overline{\Lambda}(\lambda,n)\right]
-i\alpha(n)\left[-2V_0^{(2)}(\lambda)\right]^{1/2}\left[1+\overline{\Omega}(\lambda,n)\right],
\label{WKB 3rd-order}
\end{equation}
where $\omega \in \mathbb{R}^{+}$, $\lambda$ is Regge frequency, $\lambda \in \mathbb{C}$, $n$  is overtone number, $n \in \mathbb{N}^+$, and $\alpha(n) \equiv n- 1/2$. In Eq.~(\ref{WKB 3rd-order}) the two factors, $\overline{\Lambda}$ and $\overline{\Omega}$ take the forms,
\begin{equation}
{\overline
\Lambda}(\lambda,n)=\frac{1}{{\left[-2V_0^{(2)}(\lambda)\right]}^{1/2}}
\left[\frac{1}{8}
\frac{V_0^{(4)}(\lambda)}{V_0^{(2)}(\lambda)}\left( \frac{1}{4}+
[\alpha(n)]^2 \right)   -\frac{1}{288}
\left(\frac{V_0^{(3)}(\lambda)}{V_0^{(2)}(\lambda)}\right)^2\left(
7+ 60 \, [\alpha(n)]^2 \right) \right],
\label{WKB_Lambda}
\end{equation}
and
\begin{eqnarray}\label{WKB_Omega}
{\overline \Omega}(\lambda,n)&=& \frac{1}{\left[-2V_0^{(2)}(\lambda)\right]}
\left[\frac{5}{6912}
\left(\frac{V_0^{(3)}(\lambda)}{V_0^{(2)}(\lambda)}\right)^4\left(
77+ 188 \, [\alpha(n)]^2 \right) \right. \nonumber \\
&&  \left. -\frac{1}{384}
\left(\frac{{[V_0^{(3)}(\lambda)]}^2V_0^{(4)}(\lambda)}{{[V_0^{(2)}(\lambda)]}^3}\right)
\left(
51+ 100 \, [\alpha(n)]^2 \right) +  \frac{1}{2304}
\left(\frac{V_0^{(4)}(\lambda)}{V_0^{(2)}(\lambda)}\right)^2 \left(
67+ 68 \, [\alpha(n)]^2 \right) \right. \nonumber \\
&& \left. +\frac{1}{288} \left(\frac{ V_0^{(3)}(\lambda)
V_0^{(5)}(\lambda)}{{[V_0^{(2)}(\lambda)]}^2}\right) \left( 19+ 28
\, [\alpha(n)]^2 \right) -\frac{1}{288}
\left(\frac{V_0^{(6)}(\lambda)}{V_0^{(2)}(\lambda)}\right)\left( 5 +
4 \, [\alpha(n)]^2 \right) \right].
\end{eqnarray}
Moreover, the notation $V_0^{(p)}(\lambda)$ stands for the derivatives of $V_{\lambda-\frac{D-3}{2}}(r)$ with respect to the tortoise coordinate at the peak of the potential,
\begin{equation}
V_0^{(p)} (\lambda) = \left.  \frac{\mathrm{d}^p}{{\mathrm{d}r_*}^p }
V_{\lambda-\frac{D-3}{2}}(r_*)\right|_{r_*={(r_*)}_0},
\label{derivative_Vr}
\end{equation}
where $V_{\lambda-\frac{D-3}{2}}(r)$ corresponds to  $V_{\ell}(r)$ that is now regarded as a function of complex $\lambda$ after the analytic continuation of $\ell$.
We shall solve Eq.~(\ref{WKB 3rd-order}) approximatively by assuming $|\lambda|\gg 1$ and $|\mathrm{Re} \lambda|\gg |\mathrm{Im} \lambda|$. Such an assumption is just the precondition $r_0 \approx r_c$ for $\ell \gg 1$ that we just proved in the above section. So we evaluate $V_0^{(p)} (\lambda)$ at $r_c$ rather than at $r_0$.  As done in Ref.~\cite{1002}, we also adopt $\eta_c$ in order to simplify the formulas of Regge frequency,
\begin{equation}
\eta_c \equiv \frac{1}{2}\sqrt{4f_{c}-2r_{c}^{2}f_{c}^{(2)}}.
\label{eta_c}
\end{equation}
Again taking Eqs.~(\ref{V_l}), (\ref{derivative_Vr}), and (\ref{eta_c}) into consideration, we compute the coefficients in Eqs.~(\ref{WKB 3rd-order}), (\ref{WKB_Lambda}), and (\ref{WKB_Omega}). These  coefficients are related to $V_0^{(p)} (\lambda)$ and put in  Appendix A.

Now it is ready to solve Eq.~(\ref{WKB 3rd-order}). By using Eqs.~(\ref{WKB_Lambda}), (\ref{WKB_Omega}), and  (\ref{derivative_Vr}) together with Appendix A, we derive the general formula of the Regge frequency of CR BHs,
\begin{equation}
\lambda_{n}(\omega) \approx \left[\frac{r_{c}^{2}}{f_{c}}
\omega^{2}
+a_{n}+2\eta_c^2[\alpha(n)]^{2}\epsilon_{n}(\omega)\right]^{1/2}+i\eta_c\alpha(n)\left[1+\epsilon_{n}(\omega)\right],
\label{Regge frequency}
\end{equation}
where
\begin{equation}
\epsilon_{n}(\omega)=\frac{b_{n}}{(r_{c}^{2}/f_{c}) \omega^{2}+a_{n}+\eta_c^2[\alpha(n)]^{2}}.
\label{Epsilon}
\end{equation}
Note that the formula has a similar form to that in Ref.~\cite{1002}, but the coefficients $a_n$ and $b_n$ are different from those in Ref.~\cite{1002}. The condition,
$|\lambda|\gg 1$ and $|\mathrm{Re} \lambda|\gg |\mathrm{Im} \lambda|$, leads to $\omega \gg 1$. Thus we can further expand $\lambda_{n}(\omega)$ in series of $\omega$,
\begin{equation}
\lambda_{n}(\omega)=\left[\frac{r_{c}}{\sqrt{f_{c}}}~\omega
+\frac{a_{n}}{(2r_{c}/\sqrt{f_{c}})}~\frac{1}{\omega}\right]
+i\eta_c\alpha(n)\left[1+\frac{b_n}{(r_{c}^{2}/f_{c})}~\frac{1}{\omega^{2}}\right]
+\underset{\omega \to +\infty}{\cal O}\left(
\frac{1}{\omega^3}\right).
\label{Regge frequency 2}
\end{equation}
Here $a_n$ and $b_n$ depend on $S_c$ and its derivatives $S_c^{(p)}$, and they encode nontrivial corrections to the Regge frequency via the scale factor. Such nontrivial corrections show the major difference between the conformally related BHs and non-conformal ones. The expressions of $a_n$ and $b_n$ are presented in Appendix B. According to Eq.~(\ref{Regge frequency 2}), the influence of $S(r)$ is efficiently suppressed in large-$\omega$ regimes. As a consequence, in high-frequency regimes, $\lambda_{n}(\omega) \approx r_c/\sqrt{f_{c}}~\omega+i\eta_c\alpha(n)$, {\em i.e.}, the Regge frequency becomes independent of the scale factor, which is consistent with the result of non-conformal BHs.

\subsection{Regge frequency: The conformally related Schwarzschild-Tangherlini black hole}
\label{sec:Regge_App}

 As an application of the result obtained in the above subsection, we give the expression of Regge frequency for conformally related Schwarzschild-Tangherlini black holes (CRST BHs) in the $D$-dimensional spacetime.

When $S(r)=1$, CRST BHs return to ST BHs. For the two types of BHs, the lapse function is of the following form,
\begin{equation}
f(r)=1-\left(\frac{r_{\rm h}}{r}\right)^{D-3},
\label{f(r)_ST BH_1}
\end{equation}
with
\begin{equation}
r_{\rm h}^{D-3}=\frac{16\pi M}{(D-2){\cal A}_{D-2}}, \qquad
{\cal A}_{D-2}=\frac{2\pi^{(D-1)/2}}{\Gamma[(D-1)/2]},
\label{f(r)_ST BH_2}
\end{equation}
where $M$ is the mass, $r_{\rm h}$ is the radius of event horizons, ${\cal A}_{D-2}$ is the area of unit sphere $S^{D-2}$, and $\Gamma[x]$ is the Gamma function. Using Eqs.~(\ref{PS1}), (\ref{eta_c}), (\ref{f(r)_ST BH_1}), and (\ref{f(r)_ST BH_2}), we compute the parameters, $r_c$ and $\eta_c$,
\begin{subequations}
\begin{eqnarray}
r_c&=&r_{{\rm h}}\left(\frac{D-1}{2}\right)^{1/(D-3)},
\label{rc_Schw}\\
\eta_c&=&\sqrt{D-3}.\label{eta_c_Schw}
\end{eqnarray}
\end{subequations}

For illustrations, we derive the Regge frequencies for the CRST BHs in 4-, 5-, and 6-dimensional spacetimes, respectively, by using Eq.~(\ref{Regge frequency 2}) together with Appendix B.
\begin{itemize}
\item $D=4$

For the 4-dimensional spacetime, considering Eqs.~(\ref{f(r)_ST BH_1}), (\ref{rc_Schw}) and (\ref{eta_c_Schw}), we have
\begin{subequations}
\begin{eqnarray}
r_{c}&=&\frac{3}{2}r_{\rm h}=3M,\\
\eta_c &=&1,\\
\frac{r_c}{\sqrt{f_c}}&=&3 \sqrt{3} M.
\end{eqnarray}
\end{subequations}
Thus, we obtain the Regge frequency from Eq.~(\ref{Regge frequency 2}) and Appendix B,
\begin{equation}
\lambda_{n}(\omega)=\left[3\sqrt{3}~M\omega+\frac{\sqrt{3}~a_n}{18~M\omega}\right]
+i\alpha(n)\left[1+\frac{b_{n}}{27~M^{2}\omega^{2}}\right]
+\underset{\omega \to +\infty}{\cal O}\left(
\frac{1}{\omega^3}\right),
\label{Regge_Schw_4D_2}
\end{equation}
with
\begin{subequations}
\begin{eqnarray}
a_n&=&-\frac{29}{216}+\frac{1}{4S_c^2}\Biggm[3 M^2 \left(S_c^{(1)}\right)^2-6 M^2 S_cS_c^{(2)}-8 M S_c S_c^{(1)}\Biggm]+\frac{5 [\alpha (n)]^2}{18},\label{Regge_Schw_4D_an}
\\
b_n& =& -\frac{371}{15552}-\frac{1}{8~S_c^4}\Biggm\{S_c^3 \biggm[ 9 M^4 S_c^{(4)}+42 M^3 S_c^{(3)}+28 M^2 S_c^{(2)}-4 M S_c^{(1)}\biggm]
\nonumber\\
&& - 3 S_c^2 \biggm[6 M^4 \left(S_c^{(2)}\right)^2+ 9 M^4 S_c^{(1)}S_c^{(3)}+32M^3 S_c^{(1)}S_c^{(2)}+8 M^2\left(S_c^{(1)}\right)^2\biggm]
\nonumber\\
&& -27 M^4 \left(S_c^{(1)}\right)^4+9 M^2 S_c \left(S_c^{(1)}\right)^2 \biggm[7 M^2 S_c^{(2)}+6 M S_c^{(1)}\biggm]\Biggm\}-\frac{305 [\alpha (n)]^2}{3888}.\label{Regge_Schw_4D_bn}
\end{eqnarray}
\end{subequations}

\item $D=5$

Similarly, for the 5-dimensional spacetime, we have
\begin{subequations}
\begin{eqnarray}
r_{c}&=&\sqrt{2}r_{\rm h}=4\sqrt{\frac{ M}{3 \pi }},\\
\eta_c &=&\sqrt{2},\\
\frac{r_c}{\sqrt{f_c}}&=&4 \sqrt{\frac{2M}{3 \pi }}.
\end{eqnarray}
\end{subequations}
The corresponding Regge frequency reads
\begin{equation}
\lambda_{n}(\omega)=\left[\sqrt{2}\,r_{c}\omega+\frac{a_{n}}{2\sqrt{2}\,r_{c}\omega}\right]
+i\sqrt{2}\,\alpha(n)\left[1+\frac{b_n}{2r_{c}^{2}\omega^{2}}\right]+\underset{\omega \to +\infty}{\cal O}\left(
\frac{1}{\omega^3}\right),
\label{Regge_Schw_5D_2}
\end{equation}
with
\begin{subequations}
\begin{eqnarray}
a_n&=&-\frac{5}{16}+\frac{1}{2 \pi  S_c^2}\Biggm[M \left(S_c^{(1)}\right)^2-4 M S_cS_c^{(2)}- 5 \sqrt{3 \pi } \sqrt{M} S_cS_c^{(1)}\Biggm]+\frac{3 [\alpha (n)]^2}{4},\label{Regge_Schw_5D_an}
\\
b_n& =& -\frac{101}{512}-\frac{1}{24 \pi ^2 S_c^4} \Biggm\{S_c^3 \biggm[16 M^2 S_c^{(4)}+ 60  \sqrt{3 \pi } M^{3/2} S_c^{(3)}+84\pi M  S_c^{(2)}-21 \sqrt{3} \pi ^{3/2} \sqrt{M} S_c^{(1)}\biggm]
\nonumber\\
&& - S_c^2 \biggm[40 M^4 S_c^{(1)}S_c^{(3)}+120\sqrt{3 \pi } M^{3/2} S_c^{(1)}  S_c^{(2)}+24 M^2 \left(S_c^{(2)}\right)^2+75\pi  M \left(S_c^{(1)}\right)^2\biggm]
\nonumber\\
&& -24 M^2 \left(S_c^{(1)}\right)^4+12 M S_c\left(S_c^{(1)}\right)^2 \biggm[5 \sqrt{3 \pi M} S_c^{(1)}+6 M S_c^{(2)}\biggm]\Biggm\}-\frac{31 [\alpha (n)]^2}{128}.\label{Regge_Schw_5D_bn}
\end{eqnarray}
\end{subequations}

\item $D=6$

As done in the above two cases, for the 6-dimensional spacetime, we have
\begin{subequations}
\begin{eqnarray}
r_{c}&=&\left(\frac{5}{2}\right)^{1/3}r_{\rm h}=\frac{\sqrt[3]{15M}}{2^{2/3}\sqrt[3]{\pi}},\\
\eta_c &=&\sqrt{3},\\
\frac{r_c}{\sqrt{f_c}}&=&\frac{5^{5/6} \sqrt[3]{M}}{2^{2/3} \sqrt[6]{3\pi^2} },
\end{eqnarray}
\end{subequations}
and write the corresponding Regge frequency,
\begin{equation}
\lambda_{n}(\omega)=\left[\sqrt{\frac{5}{3}}r_{c}\omega+\frac{a_{n}}{2\sqrt{5/3}~r_{c}\omega}\right]
+i\sqrt{3}\,\alpha(n)\left[1+\frac{b_{n}}{(5/3)r_{c}^{2} \, \omega^{2}}\right]+\underset{\omega \to +\infty}{\cal O}\left(
\frac{1}{\omega^3}\right),
\label{Regge_Schw_6D_2}
\end{equation}
with
\begin{subequations}
\begin{eqnarray}
a_n&=&-\frac{31}{60}-\frac{3 }{20 \pi ^{2/3} S_c}\biggm[(30M)^{2/3} S_c^{(2)}+12 \sqrt[3]{30 \pi M} S_c^{(1)}\biggm]+\frac{7 [\alpha (n)]^2}{5},\label{Regge_Schw_6D_an}
\\
b_n &=& -\frac{409}{720}-\frac{3}{800 \pi ^{4/3} S_c^3}\Biggm\{S_c^2 \biggm[15 \sqrt[3]{30} M^{4/3} S_c^{(4)}+480 \sqrt[3]{\pi } M S_c^{(3)}
\nonumber\\
&& +56 (30 \pi )^{2/3} M^{2/3} S_c^{(2)}-192 \sqrt[3]{30} \pi \sqrt[3]{M} S_c^{(1)}\biggm]
\nonumber\\
&& - S_c \biggm[15 \sqrt[3]{30} M \left(S_c^{(2)}\right)^2+30 \sqrt[3]{30} M S_c^{(1)}S_c^{(3)}
\nonumber\\
&&+840 \sqrt[3]{\pi } M^{2/3}S_c^{(1)} S_c^{(2)}+56 (30 \pi )^{2/3} \sqrt[3]{M} \left(S_c^{(1)}\right)^2\biggm]
\nonumber\\
&& +30 M^{2/3}\left(S_c^{(1)}\right)^2\biggm[12 \sqrt[3]{\pi } M^{1/3} S_c^{(1)}+\sqrt[3]{30} M^{2/3}  S_c^{(2)}\biggm]\Biggm\} -\frac{91 [\alpha (n)]^2}{180}.\label{Regge_Schw_6D_bn}
\end{eqnarray}
\end{subequations}
\end{itemize}

From the above Regge frequencies of CRST BHs, we find that $a_n$ and $b_n$ depend on $M$, $S_c$, and $S_c^{(p)}$ in a subtle way. In the second terms of $a_n$ and $b_n$, the combination of $M$, $S_c$ and $S_c^{(p)}$ is balanced such that the dimensions of all subterms in these second terms offset precisely with one another, which makes $a_n$ and $b_n$ exactly dimensionless. The powers of $M$ in $a_n$ and $b_n$ are not uniform in different  dimensional spacetimes because the dimension of mass changes with spacetime dimensions as shown in Eq.~(\ref{f(r)_ST BH_2}).

\section{Absorption cross section}
\label{sec:ACS}
Based on the Regge frequency performed above, we are now ready to compute the ACSs for CR BHs. This section is divided into  two subsections, where one focuses on the analytic expression of ACSs and the other to the illustration of ACSs in the models of CRST BHs.

\subsection{Analytic expression of absorption cross sections}
\label{abs_expressions}

As shown in Refs.~\cite{1101,Harris}, the ACS of a $D$-dimensional static spherically symmetric BH reads\footnote{This formula is valid for any effective potentials that are the function of radial coordinate only, so it is also valid in the CR BHs with a nontrivial scale factor. See also Ref.~\cite{Gubser}.}
\begin{equation}
\sigma_\mathrm{abs}(\omega)=\frac{\pi^{(D-2)/2}}{\Gamma
\left[(D-2)/2\right]
\omega^{D-2}} \sum_{\ell=0}^{+\infty} \frac{\Gamma[\ell+D-4]}{\Gamma[\ell]}
\left(2\ell + D-3\right) \Gamma_\ell(\omega),
\label{sigma_abs_D-dim}
\end{equation}
where $\Gamma_{\ell}(\omega)$ denotes the greybody factor.\footnote{Note to distinguish this factor from the Gamma function, $\Gamma[x]$.} Further, based on a modified version of Sommerfeld-Watson
transformation~\cite{Newton}, $\sigma_{\mathrm{abs}}$ can be analytically continued~\cite{1101} such that $\ell+(D-3)/2 \equiv \lambda \in \mathbb{C}$. After these performances, one can rewrite $\sigma_{\mathrm{abs}}$ which consists of three parts,
\begin{eqnarray}
\sigma_\mathrm{abs}(\omega)&=&\frac{2
\pi^{(D-2)/2}}{\Gamma\left[(D-2)/2\right] \omega^{D-2}}
\int_0^{+\infty} \frac{\Gamma[\lambda+(D-3)/2]}{\Gamma
[\lambda-(D-5)/2]}\,\lambda \Gamma_{\lambda-\frac{D-3}{2}}(\omega)
\,\mathrm{d}\lambda \nonumber\\
&&- \frac{4 \pi^{D/2}}{\Gamma\left[(D-2)/2\right]
\omega^{D-2}}\,\mathrm{Re} \Biggm[ \sum_{n=1}^{+\infty}
\frac{\Gamma[\lambda_n(\omega)+(D-3)/2]}{\Gamma
[\lambda_n(\omega)-(D-5)/2]} \frac{
e^{i\pi[\lambda_n(\omega)-(D-3)/2]}\,\lambda_n(\omega)
\gamma_n(\omega)
}{\sin[\pi (\lambda_n(\omega)-(D-3)/2)]} \Biggm] \nonumber\\
&&+ \frac{\pi^{(D-2)/2}}{\Gamma\left[(D-2)/2\right]
\omega^{D-2}} \int_0^{+i\infty} \Biggm[ i\,
\frac{\Gamma[\lambda+(D-3)/2]}{\Gamma [\lambda-(D-5)/2]} \frac{e^{i
\pi[\lambda -(D-3)/2]}\,\lambda
\Gamma_{\lambda-\frac{D-3}{2}}(\omega)}{\sin [\pi (\lambda -(D-3)/2)]}
 \nonumber\\
&&+ i\,
\frac{\Gamma[-\lambda+(D-3)/2]}{\Gamma [-\lambda-(D-5)/2]}
\frac{e^{i \pi[\lambda + (D-3)/2]}\,\lambda \Gamma_{-\lambda-\frac{D-3}{2}}(\omega)}{\sin [\pi (\lambda +(D-3)/2)]} \Biggm] \mathrm{d}\lambda,
\label{sigma_abs_explicit}
\end{eqnarray}
where $\gamma_n(\omega)$ is the residue of greybody factors at the $n$-th Regge pole,
\begin{equation}
\gamma_n(\omega) \equiv\left. {\mathrm{Res}\left[
\Gamma_{\lambda-\frac{D-3}{2}}(\omega)\right]}\right|_{\lambda=\lambda_n(\omega)}.\label{gbf}
\end{equation}
The three terms in Eq.~(\ref{sigma_abs_explicit}) play different roles. Let us give a detailed explanation.

The first term corresponds to the geometric absorption cross section, $\sigma_{\rm geo}$, which has the following form~\cite{1101},
\begin{equation}
\sigma_{\rm geo}=\frac{\pi^{(D-2)/2} b_c^{D-2}}{\Gamma[D/2]},
\label{general sigma_geo}
\end{equation}
where $b_c\equiv r_c/\sqrt{f_c}$  is independent of $S(r)$.
As a result, $\sigma_{\rm geo}$ is independent of $S(r)$, which indicates that the first term of Eq.~(\ref{sigma_abs_explicit}) is out of our central concern.

We shall see that
the third term of Eq.~(\ref{sigma_abs_explicit}) is also out of our central concern. The reason is as follows.
The integration with respect to $\lambda$ leads to an overall factor. By definition, the greybody factor is the modulus square of transmission coefficients, so it is not larger than $1$, namely, $\Gamma_{\lambda-\frac{D-3}{2}}(\omega) \leq 1$. Therefore, we know that the overall factor has an upper bound, which can be obtained by setting $\Gamma_{\lambda-\frac{D-3}{2}}(\omega)=1$ in the third term of Eq.~(\ref{sigma_abs_explicit}),
\begin{eqnarray}
\int_0^{+i\infty} \left\{ i\,
\frac{\Gamma[\lambda+\frac{D-3}{2}]}{\Gamma [\lambda-\frac{D-5}{2}]} \frac{e^{i
\pi(\lambda -\frac{D-3}{2})}\,\lambda}{\sin [\pi (\lambda -\frac{D-3}{2})]}
+ i\,
\frac{\Gamma[-\lambda+\frac{D-3}{2}]}{\Gamma [-\lambda-\frac{D-5}{2}]}
\frac{e^{i \pi(\lambda + \frac{D-3}{2})}\,\lambda }{\sin [\pi (\lambda +\frac{D-3}{2})]} \right\} \mathrm{d}\lambda.
\label{upper-bound}
\end{eqnarray}
By means of numerical integration of Eq.~(\ref{upper-bound}), we find that the effect of the third term of Eq.~(\ref{sigma_abs_explicit}) is quite small if compared to that of the second term. In fact, within the intermediate frequency region that we concern most, the third term just contributes a minor correction to the total ACS. This will be further illustrated intuitively in Sec.~\ref{sec:abs_plot}.

The second term of Eq.~(\ref{sigma_abs_explicit}) represents the contribution from the Regge poles. It manifests the oscillation behavior of ACSs with respect to $\omega$, therefore it does contain the characteristic information of CR BHs.
For this reason, the second term is our concern.

Now let us introduce the method we shall use for calculation of the second term of Eq.~(\ref{sigma_abs_explicit}). Gamma functions have the property,
\begin{equation}
\frac{\Gamma(z+a)}
{\Gamma(z+b)} \ \sim \ \left(\frac{1}{z}\right)^{-a+b},\qquad \mathrm{if} \mspace{20mu} |z| \rightarrow +\infty \mspace{20mu} \mathrm{and} \mspace{20mu} |\mathrm{arg} \ z|< \pi.
\label{Gamma_func_1}
\end{equation}
Moreover, the sine function can approximately be reduced to
\begin{eqnarray}
& &\sin\left(\pi
[\lambda_n(\omega)-(D-3)/2]\right)
\nonumber\\
& =& \ \frac{1}{2i}\left[e^{i \pi
[\lambda_n(\omega)-(D-3)/2]}-e^{-i \pi
[\lambda_n(\omega)-(D-3)/2]}\right]
\nonumber\\
& =& \ \frac{1}{2i}\left[e^{\pi \mathrm{Im}\lambda_n(\omega)}\left(e^{i\pi [\mathrm{Re}\lambda_n(\omega)-(D-3)/2]}e^{-2\pi\mathrm{Im}\lambda_n(\omega)}-e^{-i\pi [\mathrm{Re}\lambda_n(\omega)-(D-3)/2]}\right)\right]
\nonumber\\
& \approx &\ -\frac{1}{2i}\left[e^{\pi \mathrm{Im}\lambda_n(\omega)}e^{-i\pi [\mathrm{Re}\lambda_n(\omega)-(D-3)/2]}\right],
\label{sin_approx}
\end{eqnarray}
where we have omitted the term, $e^{i\pi [\mathrm{Re}\lambda_n(\omega)-(D-3)/2]}e^{-2\pi\mathrm{Im}\lambda_n(\omega)}$, because it is greatly suppressed by the factor $e^{-2\pi\mathrm{Im}\lambda_n(\omega)}$. It is obvious to see that the leading order of $e^{-2\pi\mathrm{Im}\lambda_n(\omega)}$ approximately equals $0.043$ when $D=4$ and $n=1$ and it is even smaller when $D>4$ and $n>1$. By using Eqs.~(\ref{Gamma_func_1}) and (\ref{sin_approx}), we obtain the second term of Eq.~(\ref{sigma_abs_explicit}) approximately as follows:
\begin{equation}
\sigma_\mathrm{abs}^\mathrm{RP}(\omega) \ \approx \
\frac{8\pi^{D/2}}{\Gamma\left[(D-2)/2\right] \omega^{D-2}}\,
\mathrm{Re}
\left[ \sum_{n=1}^{+\infty} ie^{2i\pi[\mathrm{Re}\lambda_n(\omega)-(D-3)/2]}
e^{-2\pi\mathrm{Im}\lambda_n(\omega)}
\left[\lambda_n(\omega)\right]^{D-3}\gamma_n(\omega) \right].
\label{sigma_RP_approx}
\end{equation}
The greybody factor can be written~\cite{1101} as
\begin{equation}
\Gamma_{\lambda-\frac{D-3}{2}}(\omega) \  = \ \left[1+\exp\left(-2\pi\frac{\omega^2-V_0(\lambda)}{\sqrt{-2V_0^{(2)}(\lambda)}}\right)\right]^{-1}.
\label{Gamma_function approx}
\end{equation}
Substituting Eq.~(\ref{Gamma_function approx}) into Eq.~(\ref{gbf}), we compute $\gamma_n(\omega)$. Further substituting $\gamma_n(\omega)$ into Eq.~(\ref{sigma_RP_approx}), we finally obtain the explicit form of $\sigma_\mathrm{abs}^\mathrm{RP}(\omega)$.

In the following we focus on the CRST BH. Although the derivations are straightforward, the residues of greybody factors in $D$-dimensional spacetimes are not listed here but just those in the four-, five-, and six-dimensional spacetimes as a sample for CRST BHs.
\begin{itemize}

\item $D=4$
\begin{eqnarray}
\gamma_n(\omega)&=&-\frac{1}{2 \pi }-i\frac{ \alpha (n)}{6 \sqrt{3} \pi  M \omega}
\nonumber\\
&&+\frac{1}{23328 \pi  M^2 S_c^4\omega ^2 }\Biggm\{486 M^3 S_c \left(S_c^{(1)}\right)^2 \biggm[7M S_c^{(2)}+6 S_c^{(1)}\biggm]-151 S_c^4-1458 M^4 \left(S_c^{(1)}\right)^4
\nonumber\\
&&-162 M^2 S_c^2 \biggm[6 M^2 \left(S_c^{(2)}\right)^2+ 9 M^2 S_c^{(1)}S_c^{(3)}+32~M S_c^{(1)} S_c^{(2)}+8 \left(S_c^{(1)}\right)^2\biggm]
\nonumber\\
&&+54 M S_c^3 \biggm[ 9 M^3 S_c^{(4)}+42M^2 S_c^{(3)}+28M S_c^{(2)}-4 S_c^{(1)}\biggm]
-276  [\alpha (n)]^2S_c^4\Biggm\}\nonumber \\
&&+\underset{\omega \to +\infty}{\cal O}\left(\frac{1}{\omega^3}\right).
\label{gamma_4D}
\end{eqnarray}

\item $D=5$
\begin{eqnarray}
\gamma_n(\omega)&=&-\frac{1}{\sqrt{2} \pi }-i\frac{ \sqrt{3} \alpha (n)}{4 \sqrt{2 \pi M} \omega }
\nonumber\\
&&+\frac{1}{2048 \sqrt{2} \pi ^2 M S_c^4\omega ^2 }\Biggm\{96 M^{3/2} S_c \left(S_c^{(1)}\right)^2 \biggm[5 \sqrt{3 \pi }  S_c^{(1)}+6 \sqrt{M} S_c^{(2)}\biggm]
\nonumber\\
&&-192 M^2 \left(S_c^{(1)}\right)^4-111 \pi ^2 S_c^4
\nonumber\\
&&-8 M S_c^2 \biggm[24 M \left(S_c^{(2)}\right)^2+40 M S_c^{(1)} S_c^{(3)}+120 \sqrt{3 \pi } \sqrt{M} S_c^{(1)} S_c^{(2)}+75 \pi  \left(S_c^{(1)}\right)^2\biggm]
\nonumber\\
&&+8 \sqrt{M} S_c^3 \biggm[16 M^{3/2} S_c^{(4)}+60 M\sqrt{3 \pi }  S_c^{(3)}+84 \sqrt{M} \pi  S_c^{(2)}-21 \sqrt{3} \pi ^{3/2} S_c^{(1)}\biggm]\Biggm\}
\nonumber\\
&&-\frac{33 [\alpha (n)]^2}{256 \sqrt{2} M \omega ^2}+\underset{\omega \to +\infty}{\cal O}\left(
\frac{1}{\omega^3}\right).
\label{gamma_5D}
\end{eqnarray}

\item $D=6$
\begin{eqnarray}
\gamma_n(\omega)&=&-\frac{\sqrt{3}}{2 \pi }-i\frac{3  \sqrt[6]{3} \alpha (n)}{\sqrt[3]{2} 5^{5/6} \pi ^{2/3} \sqrt[3]{M} \omega }
\nonumber\\
&&+\frac{1}{20000\sqrt{3} \pi ^{5/3} M^{2/3} S_c^3\omega ^2 } \Biggm\{270 M\left(S_c^{(1)}\right)^2 \biggm[12 \sqrt[3]{30 \pi }  \left(S_c^{(1)}\right)+30^{2/3} M^{1/3}  S_c^{(2)}\biggm]
\nonumber\\
&&-1468 \sqrt[3]{30} \pi ^{4/3} S_c^3-135 M^{\frac{2}{3}} S_c \biggm[\left(30M\right)^{\frac{2}{3}} \left(S_c^{(2)}\right)^2+2 \left(30M\right)^{{2}/{3}} S_c^{(1)} S_c^{(3)}
\nonumber\\
&&+56 \sqrt[3]{30 \pi } M^{\frac{1}{3}} S_c^{(1)} S_c^{(2)}+112 \pi^{\frac{2}{3}}  \left(S_c^{(1)}\right)^2\biggm]
+27 \sqrt[3]{M} S_c^2 \biggm[5\left(30^{2/3}\right) M S_c^{(4)}
\nonumber\\
&&+160 \sqrt[3]{30 \pi } M^{2/3} S_c^{(3)}+560 \pi ^{2/3} M^{1/3} S_c^{(2)}-64 \left(30^{2/3}\right) \pi S_c^{(1)}\biggm]\Biggm\}
\nonumber\\
&&-\frac{11\left(3^{5/6}\right) \sqrt[3]{2} [\alpha (n)]^2}{25\left(5^{2/3}\right) \sqrt[3]{\pi }M^{2/3} \omega ^2}+\underset{\omega \to +\infty}{\cal O}\left(
\frac{1}{\omega^3}\right).
\label{gamma_6D}
\end{eqnarray}

\end{itemize}

In the final part of this subsection, we derive the analytic expressions of $\sigma_\mathrm{abs}^\mathrm{RP}(\omega)$ for the CRST BHs. We omit all contributions from the Regge poles of higher excitations, {\em i.e.}, those with $n>1$. The reason is same as that explained above, that is,  all these contributions are suppressed by the exponential function $e^{-2\pi\mathrm{Im}\lambda_n(\omega)}$, where $\mathrm{Im}\,\lambda_n(\omega)=\eta_c \alpha(n)+\mathcal{O}(1/\omega^2)=\eta_c(n-1/2)+\mathcal{O}(1/\omega^2)$; see Eq.~(\ref{Regge frequency 2}). Here we list the results in the four-, five-,  and six-dimensional spacetimes.
\begin{itemize}
\item  $D=4$
\begin{eqnarray}
\sigma_\mathrm{abs}^\mathrm{RP}(\omega)&=&-8 e^{-\pi }  \Biggm[
\frac{3 \sqrt{3} M \pi \sin \left(6 \sqrt{3} \pi  M \omega \right)}{2 \omega}-\frac{\pi\cos \left(6 \sqrt{3} \pi  M \omega \right)}{2 \omega ^2}
\nonumber\\
&& -\frac{\pi\sin \left(6 \sqrt{3} \pi  M \omega \right) }{1296 \sqrt{3} M S_c^2 \omega^3}\left(162 M^2 S_c S_c^{(2)}-81 M^2 \left(S_c^{(1)}\right)^2+216 M S_c S_c^{(1)}+61 S_c^2\right)\Biggm]
\nonumber\\
&&+\underset{\omega \to +\infty}{\cal O}\left(
\frac{1}{\omega^4}\right).
\label{sigma_RP_Schw_4D}
\end{eqnarray}

\item  $D=5$
\begin{eqnarray}
\sigma_\mathrm{abs}^\mathrm{RP}(\omega)&=& 8 e^{-\sqrt{2} \pi } \Biggm[
\frac{32 \sqrt{2} M  \sin \left[\left(8 \sqrt{2 \pi M / 3}\right) \omega \right]}{3 \pi ^2 \omega}-\frac{4 \sqrt{6M} \cos \left[\left(8 \sqrt{2 \pi M / 3}\right) \omega \right]}{ \pi ^{3/2}\omega ^2}
\nonumber\\
&& -\frac{\sin \left[\left(8 \sqrt{2 \pi M / 3}\right) \omega\right] }{4 \sqrt{2} \pi ^2 S_c^2 \omega^3}\left(16 M S_c S_c^{(2)}-4 M \left(S_c^{(1)}\right)^2+20 \sqrt{3 \pi M} S_c S_c^{(1)}+13 \pi  S_c^2\right)\Biggm]
\nonumber\\
&&+\underset{\omega \to +\infty}{\cal O}\left(\frac{1}{\omega^4}\right).
\label{sigma_RP_Schw_5D}
\end{eqnarray}

\item  $D=6$
\begin{eqnarray}
\sigma_\mathrm{abs}^\mathrm{RP}(\omega)&=& -8 e^{-\sqrt{3} \pi } \Biggm[\frac{25 \sqrt{5} M \sin \left[\left( 5^{5/6} \pi ^{2/3} \sqrt[3]{2M}/\sqrt[6]{3}\right) \omega\right]}{8 \pi ^2 \omega}
\nonumber\\
&&-\frac{5(15^{2/3}) M^{2/3} \cos \left[\left( 5^{5/6} \pi ^{2/3}\sqrt[3]{2M} /\sqrt[6]{3}\right) \omega\right]}{2 \sqrt[3]{2} \pi ^{5/3} \omega ^2}
\nonumber\\
&&-\frac{\sqrt[3]{3M}  \sin \left[\left( 5^{5/6} \pi ^{2/3}\sqrt[3]{2M} /\sqrt[6]{3}\right) \omega\right] }{16 \sqrt[6]{5} \pi ^2 S_c\omega ^3}\left(9(15M)^{2/3} S_c^{(2)}\right.
\nonumber\\
&&\left.+54 \sqrt[3]{60 \pi M} S_c^{(1)}+95 \sqrt[3]{2} \pi ^{2/3} S_c\right)\Biggm]+\underset{\omega \to +\infty}{\cal O}\left(\frac{1}{\omega^4}\right).
\label{sigma_RP_Schw_6D}
\end{eqnarray}

\end{itemize}

\subsection{Graphs of $\sigma_\mathrm{abs}^\mathrm{RP}(\omega)$ for CRST BHs}
\label{sec:abs_plot}

Now we present the above results more intuitively in graphs. We restrict our attention onto the (CR)ST BHs and plot the behavior of $\sigma_\mathrm{abs}^\mathrm{RP}(\omega)$ with respect to $\omega$. The dimensionless quantities,  $\sigma_\mathrm{abs}^\mathrm{RP}(\omega)/ A_{{\rm h}}$ and $r_{{\rm h}}\omega$, are adopted, where $A_{{\rm h}} \equiv \mathcal{A}_{D-2}r_{{\rm h}}^{D-2}$ is the area of event horizon of a $(D-2)$-dimensional sphere. For simplicity, we refer to $\sigma_\mathrm{abs}^\mathrm{RP}(\omega)$ as the ``oscillation" in the context.

\begin{figure}[htpb]
	\centering
    \includegraphics[width=0.49\textwidth]{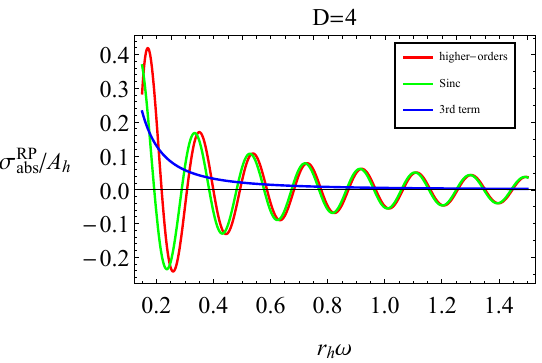}
		\vspace{5mm}
	\includegraphics[width=0.49\textwidth]{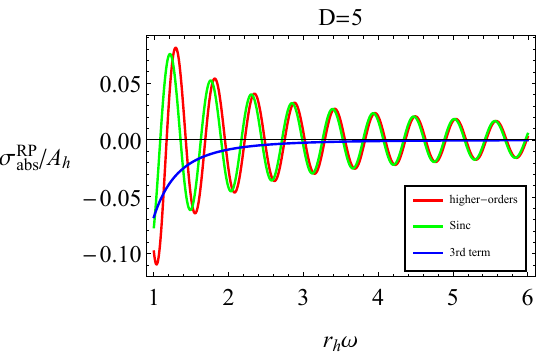}
		\vspace{5mm}
	\includegraphics[width=0.49\textwidth]{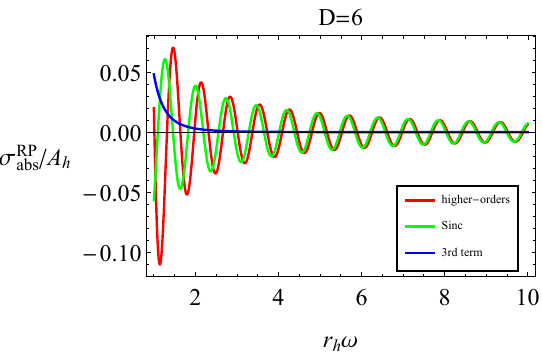}
		\vspace{5mm}
	\caption{The red curves stand for our results, the green curves for the Sinc ones in Ref.~\cite{1101}, and the blue curves for the upper bound of the third term in Eq.~(\ref{sigma_abs_explicit}).}
\label{fig:Sinc-Approx}
\end{figure}

At first, in order to justify the correctness of our results, we show that our results are consistent with the Sinc results obtained in Ref.~\cite{1101}, where the latter is based on the model of non-conformal Schwarzschild-Tangherlini BHs. Thus we set $S(r)=1$ and choose the lapse function to be that of the Schwarzschild-Tangherlini BH in our results. We plot the comparison in Fig.~\ref{fig:Sinc-Approx}, where the red and green curves  agree with each other in a wide range. Only when the frequency is  extremely low, does the disagreement appear slightly. The reason is that our results include  $\mathcal{O}(1/\omega^3)$ corrections to $\sigma_\mathrm{abs}^\mathrm{RP}(\omega)$, while the Sinc results only include the order of $\mathcal{O}(1/\omega)$. Fig.~\ref{fig:Sinc-Approx} manifests the consistency of our results with the Sinc's. Moreover, this figure also justifies that the third term of Eq.~(\ref{sigma_abs_explicit}) indeed contributes a minor correction to the total ACS; see the blue curves. The effect of blue curves decreases rapidly when the frequency increases, such that it can be omitted in the intermediate region of frequency.

\begin{figure}[htpb]
	\centering
	\includegraphics[width=0.49\textwidth]{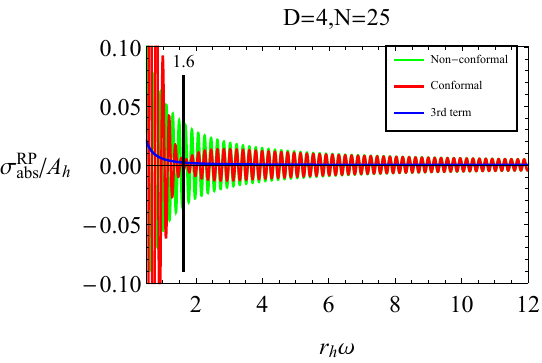}
	\vspace{5mm}
	\includegraphics[width=0.49\textwidth]{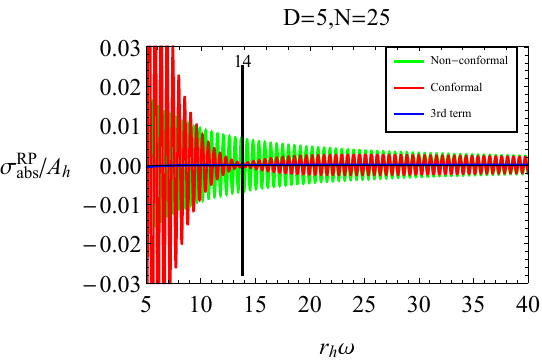}
	\vspace{5mm}
	\includegraphics[width=0.49\textwidth]{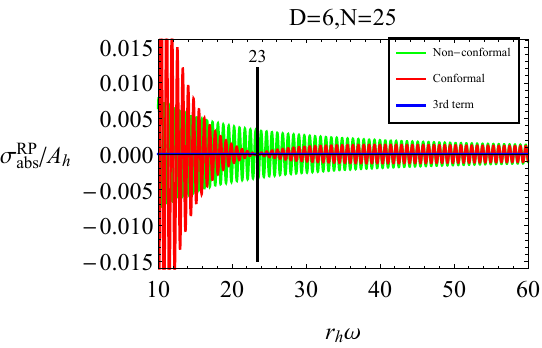}
	\vspace{5mm}
	\caption{Comparison between CRST BHs and ST BHs, where $L=r_c$ and $N=25$ are chosen for the convenience of plotting. The red curves stand for the CRST BHs and the green curves for the ST BHs, where the vertical black lines mark the positions of ``the minimum amplitude" in the four-, five-,  and six-dimensional spacetimes, respectively. The blue curves stand for the upper bound of the third term in Eq.~(\ref{sigma_abs_explicit}). Note that the blue curves are very close to the horizontal axes.}
	\label{fig:NC-C_456D}
\end{figure}

\begin{figure}[h!]
	\centering
	\includegraphics[width=0.49\textwidth]{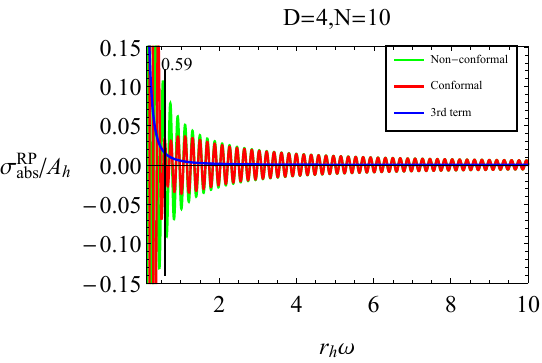}
	\vspace{5mm}
	\includegraphics[width=0.49\textwidth]{NC_C_4D_N25.pdf}
	\vspace{5mm}
	\includegraphics[width=0.49\textwidth]{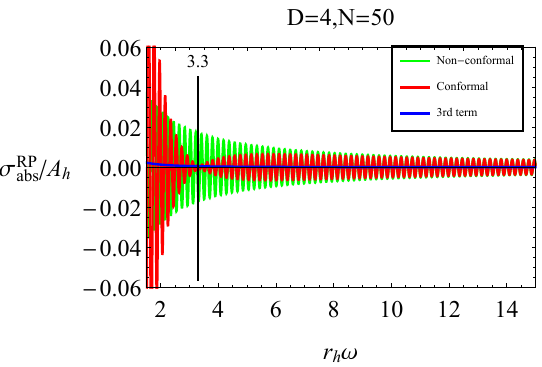}
	\vspace{5mm}	
	\includegraphics[width=0.49\textwidth]{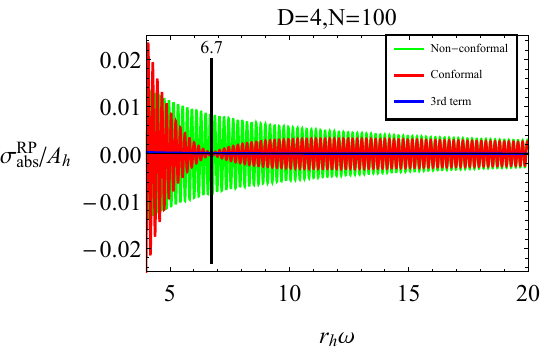}
	\caption{Comparison between the CRST BHs and ST BHs in the four-dimensional spacetime, where $L=r_c$ and $N=10, 25, 50, 100$ are set. The red curves stand for the CRST BHs and the green curves for the ST BHs, where the vertical black lines mark the positions of ``the minimum amplitude" in the cases of $N=10, 25, 50, 100$, respectively. The blue curves stand for the upper bound of the third term in Eq.~(\ref{sigma_abs_explicit}). Note that the blue curves are very close to the horizontal axes.}
	\label{fig:NC-C_4D_varyN}
\end{figure}

Next,  we investigate the influence of scale factor $S(r)$ on $\sigma_\mathrm{abs}^\mathrm{RP}(\omega)$. This will be accomplished by studying the distinction between CRST BHs and ST BHs. We consider one specific kind of scale factors,
\begin{equation}
S(r)=1+\left(\frac{L}{r}\right)^{2N},
\label{S(r)}
\end{equation}
where $N$ is a dimensionless constant and $L$ is a typical length scale of CRST BHs, {\em e.g.}, $r_{\rm h}$, or $r_c$, or the Planck length~\cite{1605}. As long as $N$ is large enough, the CRST BHs are free of spacetime singularity. In fact, we can verify that the Ricci scalar behaves as $r^{2N-D+1}$ and the Kretschmann scalar as $r^{4N-2D+2}$ when $r \rightarrow 0$ for the choice of Eq.~(\ref{S(r)}). If we take $N=25$ in the four-, five-,  and six-dimensional spacetimes in Fig.~\ref{fig:NC-C_456D} and $N=10, 25, 50, 100$ in the four-dimensional spacetime in Fig.~\ref{fig:NC-C_4D_varyN}, the corresponding CRST BHs are regular.

We present the behavior of $\sigma_\mathrm{abs}^\mathrm{RP}(\omega)$ with respect to $\omega$ in Fig.~\ref{fig:NC-C_456D} by using Eqs.~(\ref{sigma_RP_Schw_4D}), (\ref{sigma_RP_Schw_5D}), and (\ref{sigma_RP_Schw_6D}),  where the case of CRST BHs with nontrivial scale factors and the case of ST BHs with the trivial scale factor ($N=0$) are plotted. It is quite interesting that the amplitude of $\sigma_\mathrm{abs}^\mathrm{RP}(\omega)$ of CRST BHs always possesses a non-monotonic behavior, unlike that of ST BHs which is monotonic. To show this phenomenon more clearly without loss of generality, we have made an appropriate choice of parameters, such as $N=25$ and $L=r_c$. For the CRST BHs, the amplitude of $\sigma_\mathrm{abs}^\mathrm{RP}(\omega)$ is quite large in extremely low frequency regimes. As $\omega$ increases, the amplitude of $\sigma_\mathrm{abs}^\mathrm{RP}(\omega)$ decreases almost to zero at a certain value of $\omega$ which is called ``the minimum amplitude", and then the amplitude increases. Comparatively speaking, the amplitude of $\sigma_\mathrm{abs}^\mathrm{RP}(\omega)$ for the ST BHs decreases monotonically. In large-$\omega$ regimes, the oscillations of the two kinds of BHs go to the same value. We note that the effect of the third term of Eq.~(\ref{sigma_abs_explicit}) can also be neglected; see the blue curves.

We further investigate the non-monotonic behavior and find that it exists universally for a wide range of parameter $N$. To show this universality, we plot the results with various values of $N$ for the four-dimensional CRST BH  in Fig.~\ref{fig:NC-C_4D_varyN}. The situations for five- and six-dimensional models are similar and thus not presented here. We note that the effect of the third term of Eq.~(\ref{sigma_abs_explicit}) can be neglected as expected.

We also find that the behavior of $\sigma_\mathrm{abs}^\mathrm{RP}(\omega)$ is not sensitive to $L$. Only when $L$ is ridiculously large, {\em e.g.} $L>2000r_c$ in the four-dimensional CRST BH with $N=25$, does the non-monotonic behavior vanish according to our numerical calculation. Note that $N$ is the exponent in $S(r)$, while $L$ is a numerator of fraction. It is natural that the influence of $N$ is more prominent than that of $L$. Furthermore, if the astrophysical data are taken into consideration, the magnitude of $L$ is not likely to be larger than the BH mass in the geometric units~\cite{1701}. As a result, the non-monotonic behavior always appears in a physically natural range of $L$ and $N$.

We end this section with a summary of our results. After some appropriate approximation, we obtain the analytic expressions of $\sigma_\mathrm{abs}^\mathrm{RP}(\omega)$ for the four-, five-, and six-dimensional CRST BHs. When CRST BHs are reduced to ST BHs, our results agree with the Sinc's~\cite{1101}. With a specific choice of the scale factor which regularizes the CRST BH spacetimes, the amplitude of $\sigma_\mathrm{abs}^\mathrm{RP}(\omega)$ of CRST BHs is non-monotonic, which gives rise to our main conclusion: The CR BHs are substantially different from their non-conformal counterparts in the aspect of ACSs.

\section{Conclusion}
\label{sec:Conclusion}

We analyze the null geodesics of CR BHs  and prove the condition, $r_0 \approx r_c$ for $\ell \gg 1$, that is, the location of  photon spheres is independent of scale factors in the large-$\ell$ limit. Then using the third-order WKB approximation together with this condition, we obtain the Regge frequency and the ACS of a massless scalar field absorbed by CR BHs. Our analyses show that the ACSs of CR BHs depend on the scale factor, which can be used to distinguish CR BHs from non-conformal BHs.

To illustrate the distinction, we investigate a specific class of CR BHs, the CRST BHs, and find that there is the non-monotonic behavior of ACS amplitudes in a wide range of parameter $N$. This phenomenon manifests the characteristic of CR BHs in the aspect of ACSs.

Significantly, it is possible to use the above mentioned phenomenon to determine observationally the scale factor of CR BHs if the CR BHs exist in our Universe. The scale factor goes to unity asymptotically, so the CR BHs and the non-conformal ones are similar at a large astronomical distance. Nevertheless, we can still distinguish these two kinds of BHs by their absorption spectra.
In Ref.~\cite{1702}, the authors concluded that a larger $N$ causes a less violation to the energy condition, which implies that a larger $N$ seems more reasonable and natural in our  world. For a larger $N$, ``the minimum amplitude" of $\sigma_\mathrm{abs}^\mathrm{RP}(\omega)$ appears in the higher-frequency regimes of absorption spectra, which is a more prominent spectral indication of the scale factor. As for astrophysical experiments, it is technically easier to observe the high-energy rays and particles than the low-energy ones. Therefore, it is feasible to distinguish a CR BH from a non-conformal BH through the observations.

It is also feasible to search for regular BHs in this way. According to the earlier studies~\cite{1408,1505,AP393,IJMPD26,2011}, the ACSs of regular NED BHs do not possess any special spectral characteristics, such as the non-monotonic behavior. Therefore, the ACSs of NED BHs are indistinguishable from those of singular BHs, such as the Reissner-Nordstrom BH. However, this is not the case for regular CR BHs. The distinction is prominent as we have shown. Therefore, we provide a feasible approach for searching for regular BHs.

Finally, we have an argument about the Hawking radiation of CR BHs. The ACSs are related to the energy flux of Hawking radiation~\cite{1402,Page1,Page2}. For example, the following formula holds for the four-dimensional Schwarzschild BH,
\begin{equation}
\frac{d^2 E(\omega)}{d\omega dt}
\propto
 \frac{\omega^{3}\sigma_\mathrm{abs}(\omega)}
{\exp(8\pi M\omega)-1},
\label{HawkingEmissionSpectrum}
\end{equation}
which implies that the radiated energy per unit time and unit frequency is proportional to the ACS of four-dimensional Schwarzschild BHs. By integrating Eq.~(\ref{HawkingEmissionSpectrum}) with respect to $t$ and $\omega$, we can obtain the loss of BH energy through the radiation, $\Delta E$. Obviously, $\Delta E$ is a function of $N$ and $L$. This implies that the loss of BH energy does depend on the scale factor $S(r)$ although the Hawking temperature is independent~\cite{1605,1611,2102} of this factor. Such an inference is surprising because it seems that there are unknown properties hidden in the thermodynamics of CR BHs. We believe that the research for the  issue will improve our understanding of CR BHs greatly.

\section*{Acknowlegement}

The authors would like to thank C. Lan for helpful discussions about the CAM method, analytic S-matrix theory and regular black holes. They would also like to thank the anonymous referee for the helpful comments that improve this work greatly.
This work was supported in part by the National Natural Science Foundation of China under Grant Nos. 11675081 and 12175108.

\section*{Appendix A: The expressions of the terms related to $V_0^{(p)} (\lambda)$}
\label{sec:Appen_A}

Here we give the asymptotic expansions of the coefficients related to $V_0^{(p)} (\lambda)$ in the limit of $|\lambda| \rightarrow +\infty$.
\begin{eqnarray}
V_0(\lambda)&=&\frac{f_c}{r_c^2}\lambda^2+\frac{f_c}{16 r_c^2 S_c^2}\Biggm\{4 (D-2) r_c S_c \left[(D-2) f_c S^{(1)}_c+f_c r_c S^{(2)}_c+2 f_c S^{(1)}_c\right]\nonumber\\
& &+(D-6)(D-2) f_c r_c^2 \left(S^{(1)}_c\right)^2+4 S_c^2 \left[(D-2) D f_c-(D-3)^2\right]\Biggm\}+\underset{|\lambda| \rightarrow
+\infty}{\cal O}\left( \frac{1}{\lambda^2}\right).
\label{V0}
\end{eqnarray}
\begin{eqnarray}
&&\left[-2V_{0}^{(2)}(\lambda)\right]^{1/2} \nonumber\\
&=& \frac{2 f_c \eta _c}{r_c^2}\lambda-
\frac{1}{16 \eta _c r_c^2 S_c^4}\Biggm\{
\Biggm[8 f_c^2 S_c^3 \left(2 (D-2) f_c r_c S_c^{(1)}-S_c \left((D-3)^2-4 (D-2) f_c\right)\right)\Biggm]
\nonumber\\
&&+2f_c r_c S_c^2 \Biggm[S_c^2 \Biggm(\frac{4 f_c }{r_c}\left(4 (D-2) f_c^{(2)} r_c^2+3 (D-3)^2\right)
\nonumber\\
&&+\frac{1}{r_c}8 (D-2) (2 d-11) f_c^2+(D-3)^2 -r_cf^{(2)}_c \Biggm)+4 (D-9)(D-2) f_c^2 r_c \left(S_c^{(1)}\right)^2
\nonumber\\
&&+ 2 (D-2) f_c r_c S_c \Biggm(\frac{2 f_c }{r_c}\left(4 (D-2) S_c^{(1)}+7 r_c S_c^{(2)}\right)+4 f_c^{(2)} r_c S_c^{(1)}\Biggm)\Biggm]
\nonumber\\
&& +f_c^2 S_c (D-2) r_c^3 S_c S_c^{(1)} \Biggm[\frac{2 f_c }{r_c}\left((5 D-46) r_c S_c^{(2)}-10 (D-2) S_c^{(1)}\right)+(D-10) f_c^{(2)} r_c S_c^{(1)}\Biggm]
\nonumber\\
&& +4f_c^2 S_c S_c^3 \Biggm[2 (D-2) (22-5 D) f_c+(D-6) (D-2) f_c^{(2)} r_c^2+(D-2)
\nonumber\\
&&+f^{(3)}_c r_c^3-3 (D-2)^2+6 (D-2)-3 \Biggm]+2f_c^2 S_c  (34-5 D) (D-2) f_c r_c^3 \left(S_c^{(1)}\right)^3
\nonumber\\
&& +2(D-2) r_c^2 f_c^2  S_c^3 \Biggm[r_c \left(S_c^{(1)} \left(2 (D-2) f_c^{(2)}+f^{(3)}_c r_c\right)+4 f_c^{(2)} r_c S_c^{(2)} \right)\nonumber\\
&&+\frac{2 f_c }{r_c}\left(r_c \left(5 (D-2) S_c^{(2)}+6 r_c S_c^{(3)}\right)-5 (D-2) S_c^{(1)}\right)\Biggm]\nonumber\\
&&+(D-2) f_c^3 \Biggm[r_c^3 S_c \left(S_c^{(1)}\right)^2 \left((24-5 (D-2)) r_c S_c^{(2)}+4 (D-2) S_c^{(1)}\right)\nonumber\\
&&+3 (D-6) r_c^4 \left(S_c^{(1)}\right)^4+12 (D-4) S_c^4 \Biggm]
\nonumber\\
&&+(D-2) r_c^2 f_c^3  S_c^2 \Biggm[r_c S_c^{(1)} \left((D-10) r_c S_c^{(3)}-6 (D-2) S_c^{(2)}\right)
+(D-8) r_c^2 \left(S_c^{(2)}\right)^2+4 (D-2) \left(S_c^{(1)}\right)^2\Biggm]\nonumber\\
&&+2 (D-2) r_c f_c^3  S_c^3 \Biggm[r_c \left((D-2) r_c S_c^{(3)}-2 (D-2) S_c^{(2)}+r_c^2 S_c^{(4)}\right)+2 (D-2) S_c^{(1)}\Biggm]\Biggm\}\frac{1}{\lambda}\nonumber\\
&&+\underset{|\lambda| \rightarrow +\infty}{\cal O}\left(
\frac{1}{\lambda^2}\right).
\label{V2}
\end{eqnarray}
\begin{eqnarray}
\frac{V_{0}^{(4)}(\lambda)}{V_{0}^{(2)}(\lambda)}&=&-\frac{f_{c}}{2\eta_c^{2}r_{c}^{2}}
\left[16f_{c}^{2}-16r_{c}^{2}f_c
f_{c}^{(2)}+4r_{c}^{3}f_{c}f_{c}^{(3)}
\left(f_{c}^{(2)}\right)^{2}+r_{c}^{4}\left(4\left(f_{c}^{(2)}\right)^{2}+f_{c}f_{c}^{(4)}\right)\right]\nonumber\\
&&+\underset{|\lambda| \rightarrow +\infty}{\cal O}\left(
\frac{1}{\lambda^2}\right).\label{V4_V2}\\
\left(\frac{V_{0}^{(3)}(\lambda)}{V_{0}^{(2)}(\lambda)}\right)^{2}&=&\frac{r_{c}^{4}f_{c}^{2}\left(f_{c}^{(3)}\right)^{2}}
{4\eta_c^{4}}+\underset{|\lambda| \rightarrow +\infty}{\cal O}\left( \frac{1}{\lambda^2}\right).
\label{V3_V2}\\
\frac{{\left[V_{0}^{(3)}(\lambda)\right]}^{2}V_{0}^{(4)}(\lambda)}{{\left[V_{0}^{(2)}(\lambda)\right]}^{3}}&=&
-\frac{r_{c}^{2}f_{c}^{3}\left({f_c}^{(3)}\right)^{2}}{8\eta_c^{6}}\left[16f_{c}^{2}-16r_{c}^{2}f_c f_{c}^{(2)}+4r_{c}^{3}f_{c}f_{c}^{(3)} \left(f_{c}^{(2)}\right)^{2}\right.\nonumber\\
&&\left.+r_c^{4}\left(4\left(f_{c}^{(2)}\right)^{2}+f_{c}f_{c}^{(4)}\right)\right]
+\underset{|\lambda| \rightarrow +\infty}{\cal O}\left( \frac{1}{\lambda^2}\right).
\label{V3V4_V2}\\
\frac{V_{0}^{(3)}(\lambda)V_{0}^{(5)}(\lambda)}{{\left[V_{0}^{(2)}(\lambda)\right]}^{2}}
&=&\frac{r_{c}^{2}f_{c}^{3}f_{c}^{(3)}}{4\eta_c^{4}}\biggm[-10f_{c}f_{c}^{(3)}+10r_{c}f_{c}f_{c}^{(4)}
+r_{c}^{2}\left(15f_{c}^{(2)}f_{c}^{(3)}+f_{c}f_{c}^{(5)}\right)\biggm]\nonumber\\
&&+\underset{|\lambda| \rightarrow +\infty}{\cal O}\left( \frac{1}{\lambda^2}\right).
\label{V3V5_V2}\\
\frac{V_{0}^{(6)}(\lambda)}{V_{0}^{(2)}(\lambda)}&=&
-\frac{f_{c}^{2}}{2\eta_c^{2}r_{c}^{4}}
\left[-272f_{c}^{3}+408r_{c}^{2}f_{c}^{2}f_{c}^{(2)}
-88r_{c}^{3}f_{c}^{2}f_{c}^{(3)}\left(f_{c}^{(2)}\right)^{2}\right.
\nonumber\\
&& \left.+r_{c}^{4}f_{c}\left(38f_{c}f_{c}^{(4)}
-204\left(f_{c}^{(2)}\right)^{2}\right)+r_{c}^{5}f_{c}\left(104f_{c}^{(2)}f_{c}^{(3)}
+18f_{c}f_{c}^{(5)}\right)
\left(f_{c}^{(2)}\right)^{2}\right. \nonumber\\
&& \left.+r_{c}^{6}\left(34\left(f_{c}^{(2)}\right)^{3}+15f_{c}\left(f_{c}^{(3)}\right)^{2}
+26f_{c}f_{c}^{(2)}f_{c}^{(4)}+f_{c}^{2}f_{c}^{(6)}\right)\right]\nonumber\\
&&+\underset{|\lambda| \rightarrow +\infty}{\cal O}\left(
\frac{1}{\lambda^2}\right).
\label{V6_V2}
\end{eqnarray}
The expansions are kept to ${\cal O}\left( 1/\lambda^2\right)$ in the limit of $|\lambda| \rightarrow +\infty$, where the scale factor only appears in $V_0(\lambda)$ and $\left[-2V_{0}^{(2)}(\lambda)\right]^{1/2}$; see Eqs.~(\ref{V0}) and (\ref{V2}).

\section*{Appendix B: The expressions of $a_n$ and $b_n$}
\label{sec:Appen_B}
We find that the terms proportional to $\left[\alpha(n)\right]^2$ do not include $S_c$ or $S_c^{(p)}$ in $a_n$ and $b_n$.
\begin{eqnarray}
a_{n}& =& -\frac{1}{1152 \eta _c^4}\Biggm \{ -72 f_c^2 \left(4 f_c-2r_c^2 f_c^{(2)} \right)-7 r_c^6f_c \left(f_c^{(3)}\right)^2  -36 r_c^3 f_c f_c^{(3)} \left(2 f_c-r_c^2 f_c^{(2)} \right)\nonumber\\
&& -18 \left(4 f_c-2 r_c^2f_c^{(2)} \right) \biggm[(D-3)^2 \left(4 f_c-2r_c^2 f_c^{(2)} \right)+ r_c^4\left(f_c^{(2)}\right)^2\biggm]\nonumber\\
&& +\frac{9 f_c }{2 S_c^2}\left(4 f_c-2 r_c^2f_c^{(2)} \right)
\biggm[(D-2) \left(4 f_c-2 r_c^2f_c^{(2)} \right)+16 r_c^2 S_c^2 f_c^{(2)} -r_c^4 S_c^2 f_c^{(4)}
\nonumber\\
&& + 4 D r_c S_c S_c^{(1)}+4 r_c^2 S_c S_c^{(2)}+(D-6) r_c^2 \left(S_c^{(1)}\right)^2+4 D S_c^2\biggm] \Biggm \}\nonumber\\
&&+\left[\alpha(n)\right]^2\frac{r_c^3f_c}{96\eta_c^{4}} \Biggm\{24f_cf_c^{(3)}+6r_cf_cf_c^{(4)}-12r_c^2f_c^{(2)}f_c^{(3)}+r_c^3\biggm[5\left(f_c^{(3)}\right)^2-3f_c^{(2)}f_c^{(4)}\biggm]\Biggm\}.
\label{a_n}
\end{eqnarray}
\begin{eqnarray}
\label{b_n}
b_n &=& \frac{1}{442368 \eta _c^{10}}\Biggm\{ 385 r_c^{12} f_c^2 \left(f_c^{(3)}\right)^4
\nonumber\\
&& +459 r_c^6 f_c \left(f_c^{(3)}\right)^2  \left(4 f_c-2 f_c^{(2)} r_c^2\right) \Biggm[4 r_c^3 f_c f_c^{(3)}+r_c^4 f_c f_c^{(4)} -16 r_c^2 f_c f_c^{(2)}+16 f_c^2+4 r_c^4 \left(f_c^{(2)}\right)^2 \Biggm]
\nonumber\\
&& +201 \left(2 f_c- r_c^2f_c^{(2)} \right)^2 \Biggm[4r_c^3 f_c f_c^{(3)}+r_c^4f_c f_c^{(4)}-16r_c^2 f_c  f_c^{(2)}+16 f_c^2+4 r_c^4 \left(f_c^{(2)}\right)^2 \Biggm]^2
\nonumber\\
&& +456 r_c^6 f_c f_c^{(3)}  \left(2 f_c- r_c^2f_c^{(2)} \right)^2 \Biggm[10 r_c f_c  f_c^{(4)}+r_c^2f_c f_c^{(5)} -10 f_c f_c^{(3)}+15 r_c^2 f_c^{(2)} f_c^{(3)}\Biggm]
\nonumber\\
&& +120\left(2 f_c- f_c^{(2)} r_c^2\right)^3 \Biggm[ 104r_c^5 f_c  f_c^{(2)} f_c^{(3)}+26r_c^6 f_c  f_c^{(2)} f_c^{(4)}-204 r_c^4 f_c \left(f_c^{(2)}\right)^2+15 r_c^6 f_c  \left(f_c^{(3)}\right)^2
\nonumber\\
&& -272 f_c^3+34 r_c^6\left(f_c^{(2)}\right)^3 + 408 r_c^2f_c^2 f_c^{(2)}- 88 r_c^3 f_c^2 f_c^{(3)}+38 r_c^4 f_c^2 f_c^{(4)}+18r_c^5 f_c^2 f_c^{(5)}+r_c^6 f_c^2 f_c^{(6)} \Biggm]
\nonumber\\
&& -\frac{864 }{S_c^4}\left(2 f_c- f_c^{(2)} r_c^2\right)^4\Biggm[-2 (D-3)^2  r_c^2 S_c^4 f_c^{(2)}
\nonumber\\
&&+f_c S_c^2 \biggm((D-10) (D-2)  r_c^4 \left(S_c^{(1)}f_c^{(2)}\right)^2
+4 S_c^2 \left((D-3)^2+(D^2-4) r_c^2 f_c^{(2)} +(D-2) r_c^3f_c^{(3)} \right)\nonumber\\
&&  +4 (D-2) r_c^3 S_c f_c^{(2)} \left((D+2) S_c^{(1)}+4  r_c S_c^{(2)}f_c^{(2)}+ r_c S_c^{(1)}f_c^{(3)}\right)\biggm)
\nonumber\\
&&  +(D-2) f_c^2\biggm((34-5 D)r_c^4 S_c \left(S_c^{(1)}\right)^2 S_c^{(2)}-6 (D-10)r_c^3 S_c \left(S_c^{(1)}\right)^3
\nonumber\\
&& +3 (D-6) r_c^4 \left(S_c^{(1)}\right)^4+4 (D-4) S_c^4\biggm)
\nonumber\\
&& +(D-2)  r_c^2 S_c^2 f_c^2 \biggm( (D-10)r_c^2 S_c^{(1)} S_c^{(3)}+4(D-20) r_c S_c^{(1)} S_c^{(2)}
\nonumber\\
&& +(D-8) r_c^2 \left(S_c^{(2)}\right)^2-8 (D+5) \left(S_c^{(1)}\right)^2 \biggm)
\nonumber\\
&&  + 2 (D-2) r_c S_c^3 f_c^2 \biggm( 8 (D-1) S_c^{(1)} +4 (2 D+3) r_c S_c^{(2)}+(D+10) r_c^2 S_c^{(3)}+r_c^3 S_c^{(4)}\biggm) \Biggm]\Biggm\}
\nonumber\\
&&+\left[\alpha(n)\right]^2\frac{r_c^3 f_c}{110592 \eta _c^{10}} \Biggm\{
 9216f_{c}^{4}f_{c}^{(3)}+13824r_{c}f_{c}^{4}f_{c}^{(4)}
+3456r_{c}^{2}f_{c}^{3}\left[-2f_{c}^{(2)}f_{c}^{(3)}+f_{c}f_{c}^{(5)}\right]
\phantom{\left(f_{c}^{(3)}\right)^{2}}
\nonumber \\
&&
+192r_{c}^{3}f_{c}^{3}\left[72\left(f_{c}^{(3)}\right)^{2}
-99f_{c}^{(2)}f_{c}^{(4)}+f_{c}f_{c}^{(6)}\right]
\nonumber \\
&& -288r_{c}^{4}f_{c}^{2}\left[12\left(f_{c}^{(2)}\right)^{2}f_{c}^{(3)}
-29f_{c}f_{c}^{(3)}f_{c}^{(4)}+18f_{c}f_{c}^{(2)}f_{c}^{(5)}\right]
\nonumber \\
&&
+12r_{c}^{5}f_{c}^{2}\left[648\left(f_{c}^{(2)}\right)^{2}f_{c}^{(4)}
+ 17f_{c}\left(f_{c}^{(4)}\right)^{2}+56f_{c}f_{c}^{(3)}f_{c}^{(5)}
-1032f_{c}^{(2)}\left(f_{c}^{(3)}\right)^{2}-24f_{c}f_{c}^{(2)}f_{c}^{(6)}
\right]
\nonumber \\
&&
+144r_{c}^{6}f_{c}\left[28\left(f_{c}^{(2)}\right)^{3}f_{c}^{(3)}
+25f_{c}\left(f_{c}^{(3)}\right)^{3}-58f_{c}f_{c}^{(2)}f_{c}^{(3)}f_{c}^{(4)}
+18f_{c}\left(f_{c}^{(2)}\right)^{2}f_{c}^{(5)}\right]
\nonumber \\
&&
+12r_{c}^{7}f_{c}\left[-36\left(f_{c}^{(2)}\right)^{3}f_{c}^{(4)}
+75f_{c}\left(f_{c}^{(3)}\right)^{2}f_{c}^{(4)}-
17f_{c}f_{c}^{(2)}\left(f_{c}^{(4)}\right)^{2}
-56f_{c}f_{c}^{(2)}f_{c}^{(3)}f_{c}^{(5)} \right.
\nonumber \\
&& \left.  +168\left(f_{c}^{(2)}\right)^{2}\left(f_{c}^{(3)}\right)^{2} +12f_{c}\left(f_{c}^{(2)}\right)^{2}f_{c}^{(6)} \right]
\nonumber\\
&&
-72r_{c}^{8}f_{c}^{(2)}\left[12\left(f_{c}^{(2)}\right)^{3}f_{c}^{(3)}
+25f_{c}\left(f_{c}^{(3)}\right)^{3}-29f_{c}f_{c}^{(2)}f_{c}^{(3)}f_{c}^{(4)}
+6f_{c}\left(f_{c}^{(2)}\right)^{2}f_{c}^{(5)}\right]
\nonumber \\
&&
+r_{c}^{9}\left[235f_{c}\left(f_{c}^{(3)}\right)^{4}-216\left(f_{c}^{(2)}\right)^{4}f_{c}^{(4)}
-450f_{c}f_{c}^{(2)}\left(f_{c}^{(3)}\right)^{2}f_{c}^{(4)}   +
51f_{c}\left(f_{c}^{(2)}\right)^{2}\left(f_{c}^{(4)}\right)^{2}\right.
\nonumber \\
&& \left.
+168f_{c}\left(f_{c}^{(2)}\right)^{2}f_{c}^{(3)}f_{c}^{(5)} +
 360\left(f_{c}^{(2)}\right)^{3}\left(f_{c}^{(3)}\right)^{2}
-24f_{c}\left(f_{c}^{(2)}\right)^{3}f_{c}^{(6)} \right]\Biggm\}.
\end{eqnarray}

\end{document}